\documentclass[dvipsnames]{article} %
\usepackage{colm2024_conference}

\usepackage{booktabs}
\usepackage{graphicx}
\usepackage{enumitem}
\usepackage{wrapfig}
\usepackage{algorithm}
\usepackage{algpseudocode}

\usepackage{microtype}
\usepackage{amsmath}
\usepackage{colortbl}
\usepackage[utf8]{inputenc}
\definecolor{lightgray}{rgb}{0.9,0.9,0.9}
\usepackage{caption}
\usepackage{subcaption}
\usepackage{setspace}
\usepackage{url}
\usepackage{multirow}
\usepackage{colortbl}
\usepackage{tabularx}
\usepackage{blindtext}
\usepackage{pgfplots}
\pgfplotsset{compat=1.18} 
\usepackage{tikz}
\usetikzlibrary{er,positioning,bayesnet}
\usepackage{makecell}
\usepackage{tipa}
\usepackage{siunitx}
\usepackage{nicefrac}
\usepackage{tocloft}
\usepackage{listings}
\usepackage[raster,skins]{tcolorbox} %
\usepackage{xltabular}
\usepackage{adjustbox}
\usepackage{xurl}
\usepackage{rotating}
\usepackage[normalem]{ulem}
\usepackage{multicol}
\usepackage{titlesec}
\usepackage{helvet}
\useunder{\uline}{\ul}{}

\usepackage{amsmath,amsfonts,bm}

\def\eqref#1{equation~\ref{#1}}

\def\1{\bm{1}}

\DeclareMathAlphabet{\mathsfit}{\encodingdefault}{\sfdefault}{m}{sl}
\SetMathAlphabet{\mathsfit}{bold}{\encodingdefault}{\sfdefault}{bx}{n}

\renewcommand{\ghlink}{https://github.com/alibaba/recis}

\newcommand*\justify{%
  \fontdimen2\font=0.4em
  \fontdimen3\font=0.2em
  \fontdimen4\font=0.1em
  \fontdimen7\font=0.1em
  \hyphenchar\font=`\-
}

\renewcommand{\texttt}[1]{%
  \begingroup
  \ttfamily
  \begingroup\lccode`~=`/\lowercase{\endgroup\def~}{/\discretionary{}{}{}}%
  \begingroup\lccode`~=`[\lowercase{\endgroup\def~}{[\discretionary{}{}{}}%
  \begingroup\lccode`~=`.\lowercase{\endgroup\def~}{.\discretionary{}{}{}}%
  \catcode`/=\active\catcode`[=\active\catcode`.=\active
  \justify\scantokens{#1\noexpand}%
  \endgroup
}

\title{RecIS: Sparse to Dense, A Unified Training Framework for Recommendation Models}

\author{
\bf RecIS Team
}

\begin{document}

\maketitle

\begin{abstract}
Search, recommendation, and advertising are core AI applications with immense business value. Since 2022, large models such as multimodal and Large Language Models have made breakthrough progress. The integration of large models with recommendation systems is rapidly changing the landscape of recommendation technology, with various new algorithms and training paradigms constantly emerging. Combining these with the PyTorch ecosystem, which is dominated by large models, has become an inevitable trend for the development of recommendation systems.

In this paper, we propose RecIS, a Unified Sparse-Dense training framework designed to achieve two primary goals:

\begin{itemize}[leftmargin=*]

\item \textbf{ Unified Framework} To create a Unified Sparse-Dense training framework based on the PyTorch ecosystem that meets the training needs of industrial-grade recommendation models that integrated with large models.

\item \textbf{ System Optimization} To optimize the sparse component, offering superior efficiency over the TensorFlow-based recommendation models. 
The dense component, meanwhile, leverages existing optimization technologies within the PyTorch ecosystem.

\end{itemize}

Currently, RecIS is being used in Alibaba for numerous large-model enhanced recommendation training tasks, and some traditional sparse models have also begun training in it.
\end{abstract}



\section{Introduction}

\subsection{Background}
Modern recommendation systems are undergoing a paradigm shift driven by the trend of \textit{scaling up}---in both data volume and computational capacity. On the data side, this entails not only extending the length of user behavior sequences but also significantly increasing the size of training samples. On the computational side, there is a growing transition from traditional Multi-Layer Perceptron (MLP) architectures to Transformer-based models, which offer superior scalability and representational power for large-scale sequential data.

\subsection{Problem Statement}

This paradigm shift has given rise to a \textbf{large-scale sparse-dense hybrid architecture} as the de facto standard in state-of-the-art recommendation systems. This framework integrates two complementary components that jointly enable high-performance modeling over massive datasets:
\begin{itemize}[leftmargin=*]
\item \textbf{Large-Scale Sparse Component:}
Contemporary recommendation models process trillions of sparse categorical features---such as user and item IDs---through embedding layers. These discrete identifiers are mapped into low-dimensional dense vectors via large embedding tables, forming the foundational representations for downstream processing.

\item \textbf{Large-Scale Dense Component:}
This component operates on embedded features using deep neural architectures with substantial model capacity---often scaling to billions of parameters. Typically based on Transformers or other highly expressive structures, this module captures complex patterns in user behavior through computationally intensive operations.
\end{itemize}

Although this hybrid architecture unlocks modeling capabilities, it also introduces critical system-level challenges. Designing a framework that simultaneously achieves scalability, efficiency, and performance across sparse and dense computation paths remains an open and pressing problem. 

Bridging the gap between these two regimes requires co-design at the algorithmic, architectural, and systems levels --- an emerging frontier in the development of next-generation recommendation platforms.

\subsection{Challenges}

Despite the growing integration of large models into recommendation systems, there is a fundamental gap in the \textbf{ Unified Sparse-Dense framework} for sparse and dense components. Existing frameworks are often optimized for only one side of this divide:

\begin{itemize}[leftmargin=*]
  \item Industrial-scale recommendation systems have long relied on \textbf{TensorFlow}, due to its mature support for large embedding tables, distributed training, and production stability. However, TensorFlow's static graph model and slower iteration cycle hinder rapid experimentation with novel dense architectures.
  
  \item In contrast, the research community increasingly favors \textbf{PyTorch} for its dynamic computation graph, rich ecosystem, and seamless integration with multimodal and large language models. However, PyTorch lacks native production-grade support for large-scale sparse modeling.

  \item We want to leverage PyTorch's extensive large-model ecosystem and technology. Additionally, we aim to capitalize on the benefits it provides, including its user-friendliness, straightforward debugging, and consistent API compatibility across updates.
\end{itemize}

So we prefer to build the \textbf{ unified Sparse-Dense framework} based on PyTorch. But the divergence above leads to challenges.

\begin{itemize}[leftmargin=*]
  \item \textbf{Modeling Challenges}: Almost all cutting-edge large models have their first official or community implementations based on PyTorch. This makes PyTorch the go-to platform for large models. However, PyTorch lacks native support for sparse modeling.
  \item \textbf{System Challenges}: For large models, the PyTorch ecosystem also has a very complete AI infrastructure, including efficient computation and distributed training technologies. However, the sparse component introduces new problems, such as IO bottlenecks, concurrency disasters, and memory bandwidth bounds. 
\end{itemize}

\subsection{Proposed Solutions}

To address these challenges, we propose a large-scale unified training framework for Sparse-Dense hybrid recommendation systems, with optimizations that span algorithmic design, system architecture, and performance modeling. We named this work RecIS, an acronym for ``\textbf{Rec}ommendation \textbf{I}ntelligent \textbf{S}ystem''.

\begin{figure}[ht]
  \centering
  \includegraphics[width=\linewidth]{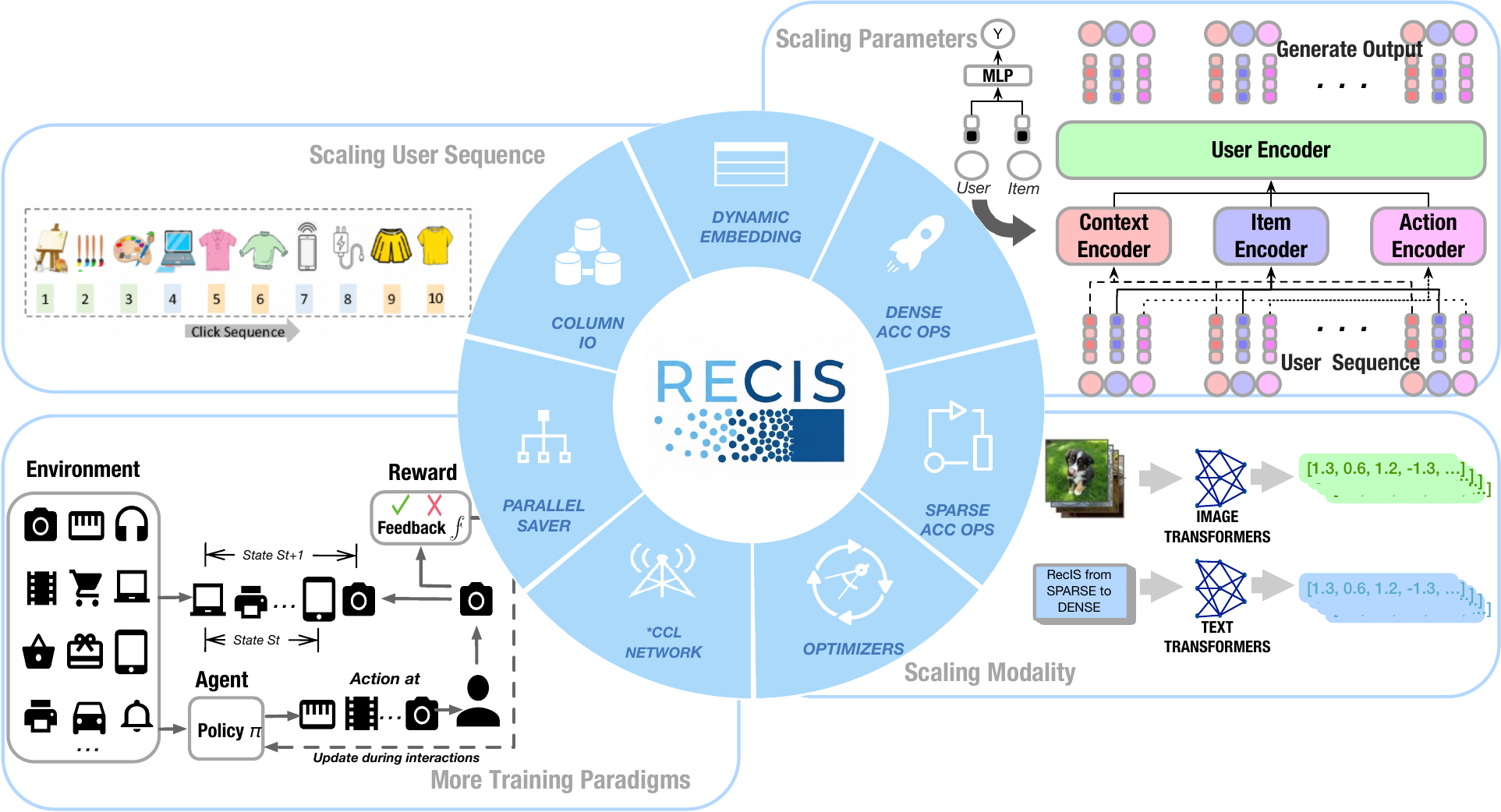}
  \caption{RecIS and Apps.}\label{fig:recis-apps}
\end{figure}

\subsubsection{Unified Sparse and Dense Modeling in PyTorch}
\begin{itemize}[leftmargin=*]
  \item \textbf{Porting Sparse Component:} RecIS must simultaneously support industrial-grade sparse training (e.g., billion-scale embeddings) and leverage the fast-evolving PyTorch ecosystem for dense modeling (e.g., large models). The sparse component is ported from the TensorFlow-based training framework to RecIS to meet the features of building large-scale sparse training. The dense component, meanwhile, is entirely based on the PyTorch ecosystem and technology itself.

  \item \textbf{Ecosystem Compatibility:} As the sparse components are ported, its operator implementations and training configurations maintain compatibility with the former system. For example, hyper-parameters and weights previously trained on the old system can be easily aligned with RecIS. 
\end{itemize}

\subsubsection{Optimizing for Sparse and Dense}
\begin{itemize}[leftmargin=*]
  \item \textbf{Performance Modeling}: 
    \begin{itemize}
      \item Performance bottlenecks for sparse and dense components differ. The dense components are typically computationally intensive, and the sparse components are classically memory bandwidth bound.

      \begin{figure}[h]
        \centering
        \begin{minipage}{0.48\textwidth}
          \centering
          \includegraphics[width=\textwidth]{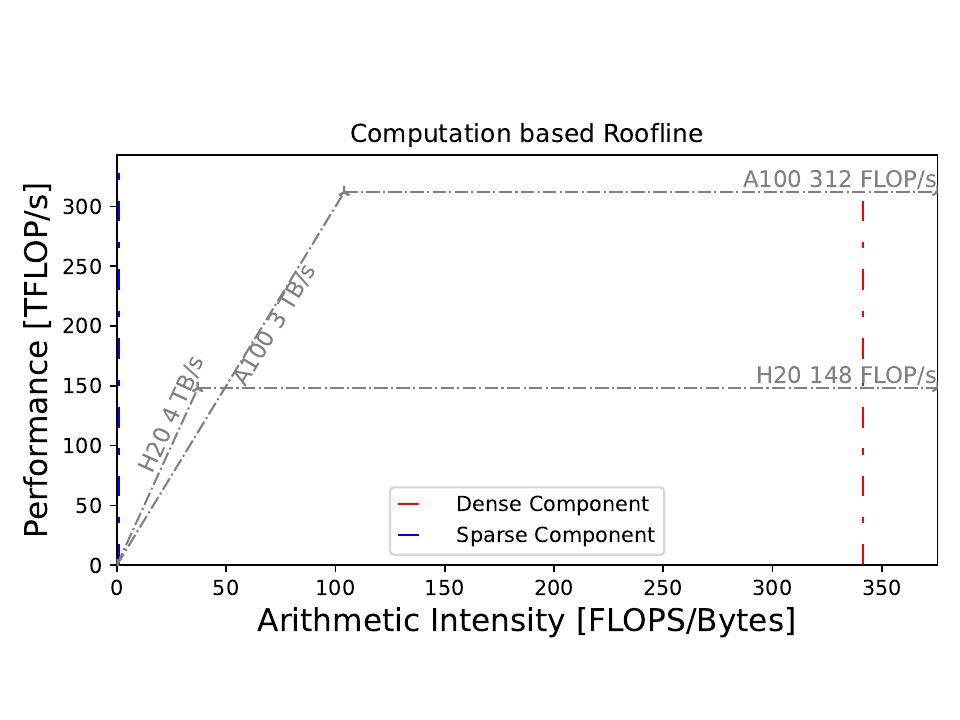}
          \caption{Computation based roofline}
          \label{fig:roofline_flops}
        \end{minipage}\hfill
        \begin{minipage}{0.48\textwidth}
          \centering
          \includegraphics[width=\textwidth]{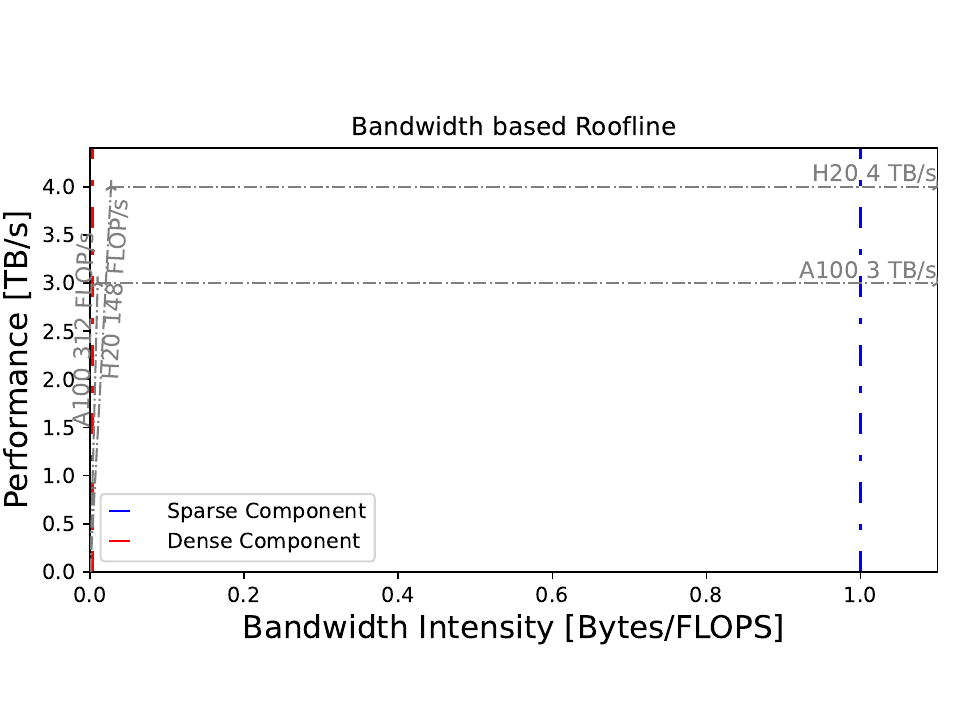}
          \caption{Bandwidth based roofline}
          \label{fig:roofline_bytes}
        \end{minipage}

\end{figure}
      
      \item The roofline model~\cite{williams2009roofline} (left) is a visual performance model used to analyze the performance of computer programs on a given hardware architecture. It focuses on Model FLOPS Utilization (MFU) and is helpful for compute-intensive workloads, such as the dense component or especially Large Language Models (LLMs). 
      
      \item However, the sparse components of the model are a classic memory-bound bottleneck. The Arithmetic Intensity (AI) of operators such as unique, embedding lookup, embedding reduce are lower than 1, which places them far to the left on the Roofline model's x-axis. Consequently, the computation-based Roofline model offers little meaningful insight for analyzing the performance of the sparse components. 
      
      \item We propose a \textbf{ bandwidth-based roofline model} (right), where the x-axis is not the ``Arithmetic Intensity'' but the ``Bandwidth Intensity'', and the y-axis is Bandwidth, which focuses on Model Bandwidth utilization (MBU). 
      In this model, sparse operators, due to their high bandwidth intensity, all reside on the roofline theoretically, which indicates that the memory-bound capacity is their optimized roofline. We should use this model to evaluate and optimize the performance of the sparse component.
    \end{itemize}
  
  \item \textbf{Performance Optimization}:
    \begin{itemize}
      \item \textbf{ Breaking through the IO Wall} Recommendation systems are highly IO intensive. To adapt to changes in sample distribution, recommendation systems need to be continuously trained on new samples. 
      This demands extremely high IO processing efficiency for the data-fetching and pre-processing work done before training. Only when IO is fully optimized can the performance bottleneck shift to the computational part.
      \item \textbf{ Breaking through the Memory Wall} The sparse large-scale component is bounded by the memory bandwidth; We take many efficient optimizations to improve the MBU to the bandwidth-based roofline.
      \item \textbf{ Breaking through the Computation Wall} The large-Scale dense component is computationally intensive. We primarily leverage large model optimization techniques from the PyTorch ecosystem to improve MFU.
    \end{itemize}
\end{itemize}

\subsection{Contributions}

Our work makes the following key contributions:

\begin{enumerate}[leftmargin=*]
  \item \textbf{A Production-Ready PyTorch based unified Sparse-Dense training framework}:
  We present RecIS that supports large-scale sparse training while seamlessly integrating with the modern large model ecosystem. 
  Compared to \textbf{TorchRec}, our framework emphasizes \textbf{industrial readiness}, with support for conflict-free embedding, efficient IO, and sparse processing.

  \item \textbf{Memory-Centric Performance Modeling}:
  We establish \textbf{MBU} as a first-class metric for recommendation systems, analogous to MFU in large models. This provides a principled way to evaluate and optimize system efficiency.

  \item \textbf{End-to-End Performance Optimization}:
  By breaking through the IO wall, the memory wall, and the computation wall, we achieve up to \textbf{ 2$\times$ higher training throughput} on large-scale recommendation tasks.

  \item \textbf{Backward Compatibility and Deployment}:
  The framework supports loading TensorFlow checkpoints and optimizers, allowing for smooth migration. It has been deployed across multiple production tasks, including search ranking and ad targeting, etc., delivering significant improvements in both efficiency and model accuracy.
  
\end{enumerate}
\section{System Design}
\subsection{Framework Components}

\begin{figure}[ht]
  \centering
  \includegraphics[width=\linewidth]{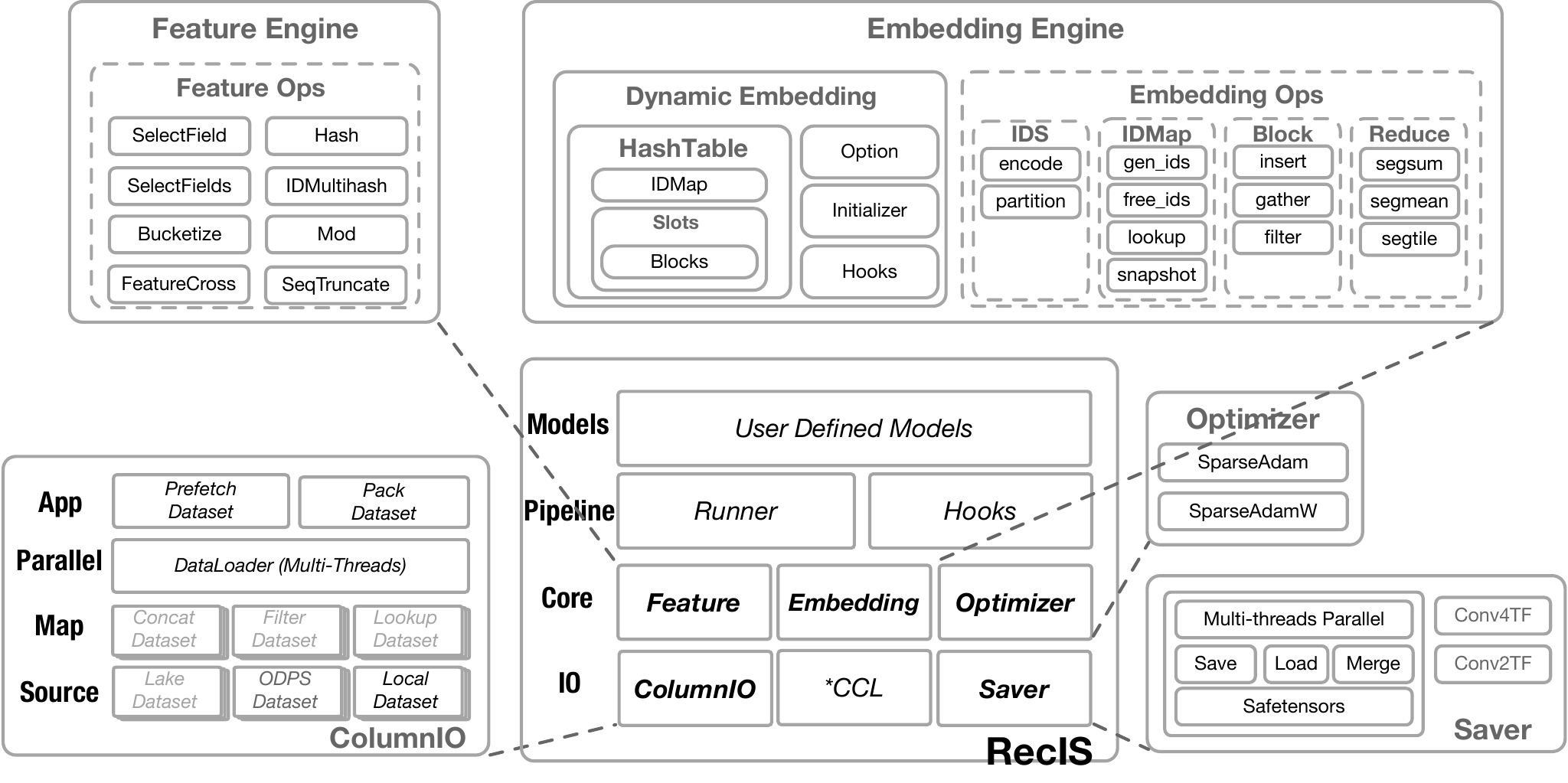}
  \caption{Framework Components.}\label{fig:recis-framework}
\end{figure}

At the framework level, to support industrial-grade sparse training and meet the demands of large-scale Embedding algorithms, we provide the following core components:
\begin{itemize}[leftmargin=*]
  \item \textbf{ColumnIO:} 
  In distributed training tasks, each GPU handles a portion of the data. The ColumnIO component efficiently reads multi-column samples from a distributed file system in a sharded manner. It assembles the samples into PyTorch Tensors for subsequent training.
    \begin{itemize}
      \item Data Type Structures: Supports types such as float, double, bigint, string and nested list which handles multi-value, and sequential features. It also supports flat features and features aggregated by a specified column (e.g., user).
      \item Data Sources: Supports batch static tables at the day-level and real-time streaming tables at the second-level. Point-lookup is also supported for joining features (e.g., Lookup multimodal features)~\cite{sheng2024enhancing}.
      \item Production Environment Support: In the Alibaba Cloud production environment, ColumnIO can directly read data from Alibaba Cloud's Distributed File System (DFS), including tables produced by MaxCompute \cite{maxcomptue}, and supports efficient reading through the tunnel \cite{tunnel}.
    \end{itemize}

  \item \textbf{Feature Engine:} 
  This component supports feature preprocessing during the training phase, which provides some flexibility for feature engineering.
    \begin{itemize}
      \item Feature transformation: Converts string features to ID types using hashing.
      \item Feature discretization: Discretize numerical features into bins.
      \item Sequence Processing: Handles operations like truncating or padding sequential features.
      \item Feature Crossing: Cross-combine ID features from different columns.
    \end{itemize}

  \item \textbf{Embedding Engine:} 
  This component provides a scalable and conflict-free embedding table using a key-value (KV) storage approach. It supports evicting stale features during continuous training and offers efficient computation and optimizer updates for sparsely stored Embeddings.
  
  \item \textbf{Saver:} 
  The model parameters associated with the sparse components will be sharded and loaded/saved in parallel using the standard SafeTensors \cite{safetensors}  format. This format will also be used for delivery to the online inference service.

  \item \textbf{Optimizer:} 
  Optimizers for Embedding, including SparseAdam and SparseAdamW, maintain compatibility with TensorFlow.

  \item \textbf{Pipelines:} 
  This component connects the aforementioned components to encapsulate the training workflow. It supports complex training processes such as multistage training (interleaving training and testing), online learning through windows, and multitask training.
\end{itemize}

\subsection{System Optimization}

\subsubsection{IO Optimization}
During the sample reading (I/O) phase, the expectation is that samples can be sufficiently supplied to the GPU for training computation, and the latency of reading samples can be hidden by the training latency. Specifically, there are multiple optimization objectives:
\begin{enumerate}[leftmargin=*]
    \item High compression rate for sample storage;
    \item Maximized throughput for reading sample data;
    \item High efficiency in processing samples.
\end{enumerate}

This primarily includes the following optimizations:

\noindent\textbf{Columnar Format:}
In a columnar storage layout, samples are stored column by column. Since the training process involves assembling various columns into multiple tensors for batch computation, columnar storage eliminates the overhead of copying data from rows during this assembly process.
Furthermore, it enables zero-cost column selection directly at the storage back-end, which satisfies feature filtering requirements of recommendation models. Additionally, columnar storage provides a better compression ratio, which not only unlocks greater capacity, but also alleviates network bandwidth pressure.

\noindent\textbf{High Concurrency and Asynchronous Operations:}
We leverage multi-threading for high concurrency reading of sharded data to maximize the I/O throughput of distributed storage. Compared to PyTorch's Python-based multi-processing DataLoader, the C++ multi-threading approach delivers higher performance and incurs lower impact on other GPU kernel launches. Through asynchronous design, we enable overlapping of I/O operations and training computations, effectively hiding the latency of data sampling.

\noindent\textbf{Memory Layout - CSR~\cite{enwiki:1300835532}:}
Compared to TensorFlow's SparseTensor(COO Format)\cite{sparse_tensor}, RaggedTensor(CSR Format)\cite{ragged_tensor} offers a more efficient storage representation, particularly for ultra-long sequence features.

\noindent\textbf{GPU Batching:}
In scenarios with extreme demands for I/O performance, we also provide a GPU batching capability.This approach both fully leverages GPU memory bandwidth and enables efficient CPU-to-GPU data copying to be pre-executed across multi-threading.

\subsubsection{Optimization for Memory Bound}
As the spare component is usually memory bound, we optimize the MBU on the critical path. This primarily involves the following core strategies:

\noindent\textbf{Moving to GPU}

In the past, some memory-intensive operators were not run on the GPU to save HBM or leverage the CPU. A heterogeneous computing architecture was taken where the model's dense computations and embedding computations were placed on the GPU, while tasks like sample parsing and embedding parameter were handled by the CPU.
This design was based on two main considerations at the time:
\begin{itemize}[leftmargin=*]
  \item HBM Savings: The massive number of embedding parameters in recommendation models consumes a significant amount of HBM. Placing them on the CPU freed up HBM for dense computations on the GPU.
  \item Heterogeneous Parallelism: This architecture leveraged the parallelism between the CPU and GPU, allowing tasks like sample reading to run in parallel with model training, or CPU-based Embedding I/O to run in parallel with GPU computations.
\end{itemize}
However, this architecture is now outdated given current hardware trends. As single-node multi-GPU setups (8 cards or more) become mainstream, the GPU's memory bandwidth is now two orders of magnitude greater than the CPU's. At the same time, GPU HBM capacity continues to increase.

\begin{figure}[ht]
  \centering
  \includegraphics[width=\linewidth]{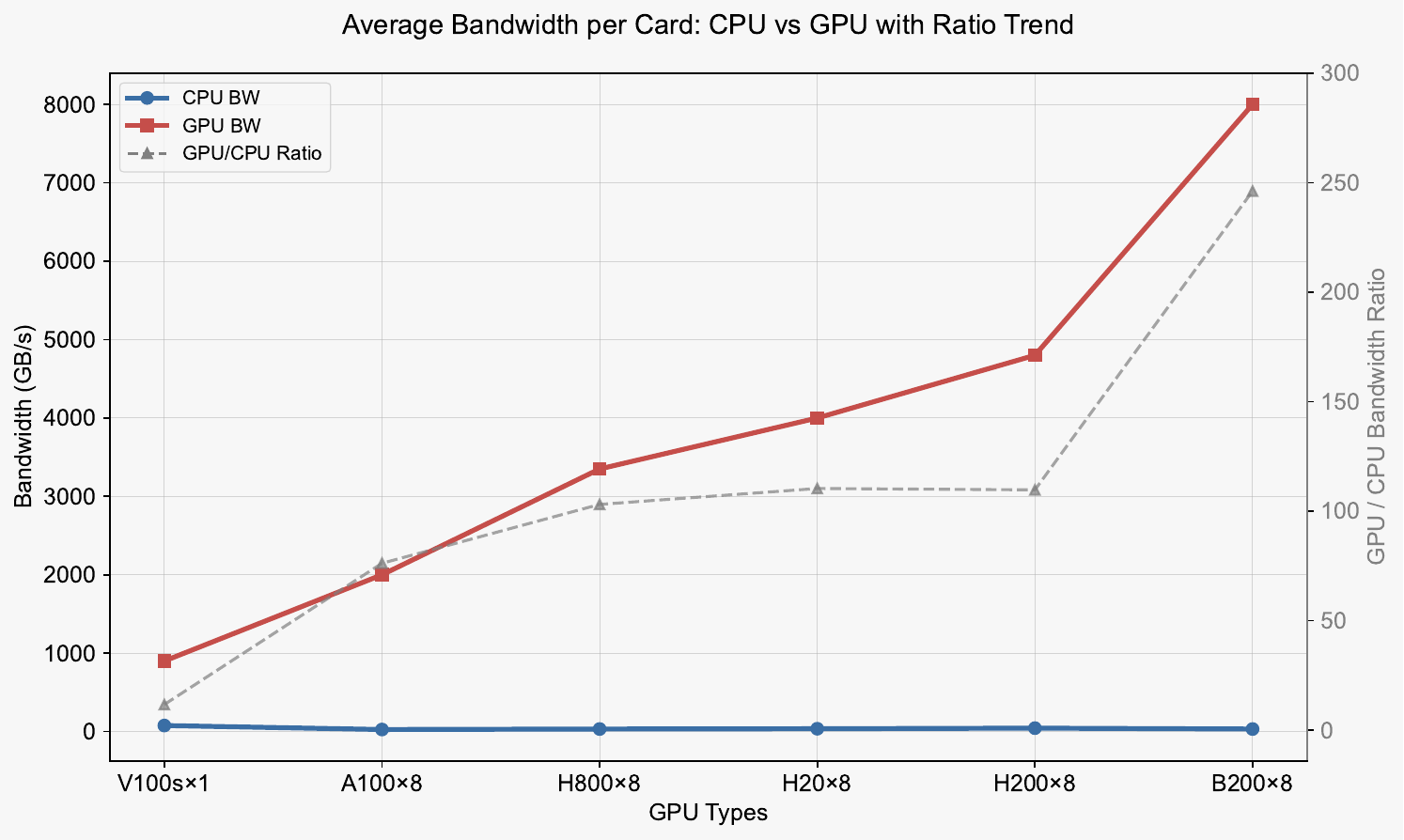}
  \caption{Due to the single-machine, multi-GPU architecture, processes on the machine share the CPU's bandwidth, the GPU's bandwidth is already two orders of magnitude greater than that of the CPU per card.}
  \label{fig:recis-bw}
\end{figure}

Therefore, to fully utilize the performance potential of this new hardware, we must shift our strategy: migrate all high-memory-demand tasks to the GPU to maximize its immense bandwidth advantage.

The RecIS framework achieves efficient dynamic embeddings through a two-tier storage architecture:
\begin{itemize}[leftmargin=*]
  \item IDMap: The first tier of storage, which uses a feature ID as the key and an offset as the value.
  \item Blocks: The second tier, a set of continuously sharded memory blocks used to store embedding parameters (e.g., float type) and their corresponding optimizer states. Blocks supports dynamic sharding for flexible scaling.
\end{itemize}

Workflow:
\begin{itemize}[leftmargin=*]
  \item Forward Pass: The model performs a single IDMap lookup to quickly find the corresponding offset in the blocks based on the feature ID. Then the embedding vector is read directly from that offset.
  \item Backward Update: During backpropagation, the system uses the computed embedding gradients and the retained forward-pass offsets to directly update the optimizer states and parameter values in the blocks.
\end{itemize}

In current multi-GPU training environments, GPU HBM is generally sufficient. In this scenario, both IDMap and Blocks can be stored in GPU HBM, fully leveraging its high bandwidth to significantly boost performance.

However, in HBM-constrained situations, either the IDMap or the Blocks can be flexibly offloaded to CPU memory to save GPU HBM. This design ensures the flexibility of the framework and its broad applicability across different hardware configurations.

\noindent\textbf{Load Balancing}

In recommendation systems, manual feature engineering is a key method for improving model effectiveness. In Alibaba's e-commerce scenarios, models often have up to thousands of feature columns, each with corresponding sparse embedding parameters. To avoid performance hot spots, these sparse parameters must be uniformly distributed in both storage space and access volume.

To achieve this, the RecIS framework employs a strategy of aggregation and full sharding:

\begin{itemize}[leftmargin=*]
  \item Parameter Aggregation \& Sharding: During model loading, we merge embedding parameters with the same dimension into a single logical table, which is then evenly distributed across multiple GPUs for storage.
  \item Request Merging \& Sharding: During the forward pass, we also merge embedding requests of the same dimension. While duplicating the requests, we also shard them based on their IDs. Subsequently, we use the All-to-All collective communication primitive to enable each card to efficiently retrieve the required Embedding vectors from other cards.
  \item Gradient Updates: During backpropagation, we also use All-to-All communication to synchronize and update the Embedding parameters and optimizer states across all cards.
\end{itemize}
The core advantage of this design is that, when the volume of merged and uniqued IDs is very large, their hash-based binning results tend toward a uniform distribution, in accordance with the Law of Large Numbers. This inherent uniformity fundamentally ensures load balancing for sparse parameters across multiple GPUs, significantly enhancing training efficiency and performance.

\noindent\textbf{Maximizing Bandwidth Utilization}

To maximize the overall performance of sparse training approaching the memory roofline, every operator on the critical path must maximize its memory access utilization. Our optimization efforts are focused on three main areas:
\begin{itemize}[leftmargin=*]
  \item \textbf{ GPU Concurrency Optimization}: Due to their sparsity and complex feature engineering, recommendation models often involve up-to thousands of variable-sized feature columns for sample input, feature processing, and embedding computation. Some columns may be empty, while others may contain up to a million feature IDs. This massive scale variation presents a dual challenge: excessive Python operations and inefficient GPU parallelism for such a large number of variable-sized kernels.
  To address this, RecIS employs a kernel fusion strategy that optimizes both horizontally and vertically. We classify operators by computation type and package adjacent fusible operators. Through this, we successfully merge thousands of small kernels in the model into just a dozen or so large ones. This not only effectively reduces launch overhead, but also significantly boosts GPU parallelism.
  
  \item \textbf{ Memory Coalescing}: Beyond coalescing sparse computations, memory accesses for Embedding can also be coalesced. In practice, we observe that the vast majority of features typically employ identical embedding dimensions (e.g., 8, 16, 32, 64, or 128). Using this characteristic, the Embedding Engine automatically performs coalescing optimizations for embeddings sharing the same dimension. This mechanism significantly reduces the number of independent memory access operations required by the model and substantially improves the efficiency of memory space utilization, thus effectively lowering overall computational and storage costs.
  
  \item \textbf{Vectorized Memory Access}: As hardware evolves, while the total HBM bandwidth increases, the bandwidth per Streaming Multiprocessor (SM) increases even more. This means that to saturate the memory bandwidth on new architectures, each SM must handle more data. For example, the H20 requires a higher "bytes-in-flight"\cite{bytes-in-flight} than the H100 to saturate memory bandwidth, a requirement that is largely on par with the B200.

  \begin{figure}[ht]
    \centering
    \includegraphics[width=\linewidth]{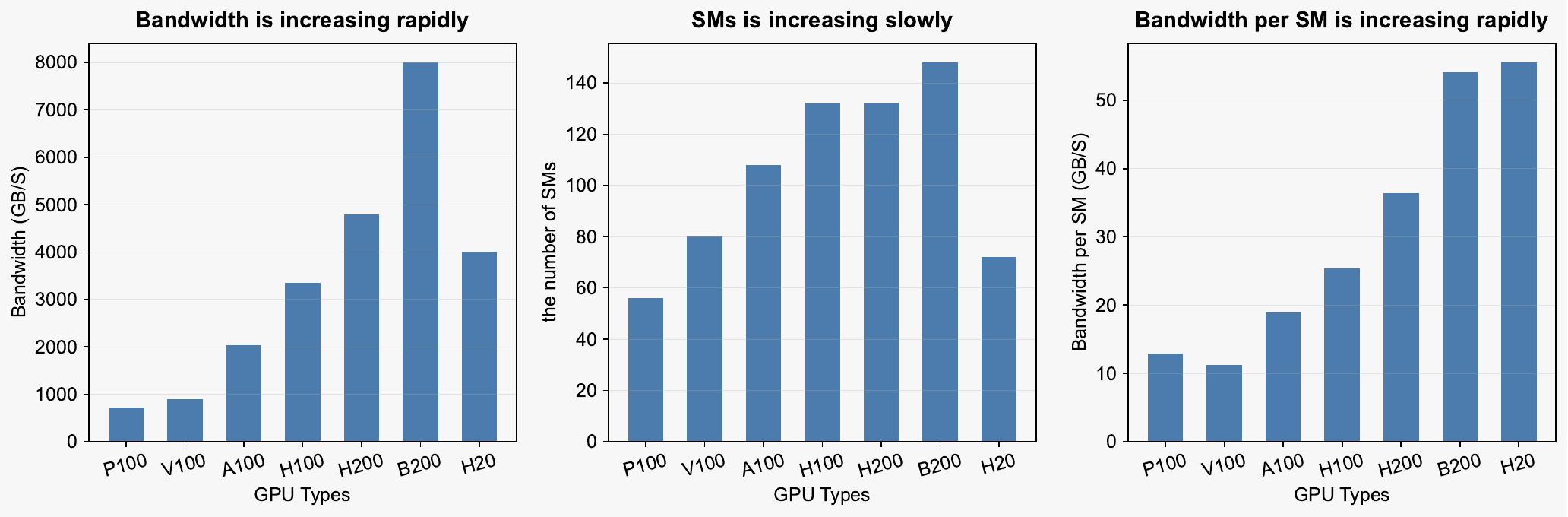}
    \caption{While the total HBM bandwidth increases, the
bandwidth per Streaming Multiprocessor (SM) increases even more. }\label{fig:bandwidth_per_sm}
  \end{figure}

  Therefore, our memory access strategy must adapt to these new hardware characteristics. Using techniques such as vectorization, we increase the parallelism of data loading, ensuring that each SM can efficiently read data from HBM and thus maximize overall memory access throughput.
  \item \textbf{Atomic Operation Optimization}: In recommendation models, reduction is a basic operation for embedding pooling. Due to the inherent variable-length nature of sparse tensors, these operations are often performed using atomic operations such as AtomicAdd, which can cause memory contention and inefficient memory access.
  To solve this, we observed that adjacent Embedding vectors in Sparse Reduction are more likely to be reduced together. Based on this insight, RecIS employs an innovative optimization: by using Warp-level merging combined with vectorized memory access, we significantly reduce atomic collisions and boost memory access efficiency.
\end{itemize}

\subsubsection{Optimization for large-model}

In training large models for recommendation, the dense component is typically the core compute-bound section. To fully unleash its performance, it is crucial to maximize the utilization of computational units.

To achieve this, we primarily leverage the mature large-model optimization techniques available in the current PyTorch ecosystem, including but not limited to:
\begin{itemize}[leftmargin=*]
  \item Mixed Precision Training \cite{micikevicius2017mixed}: We typically use FP32 for the sparse computation part and FP16/BF16 for the dense attention part. This significantly accelerates training and reduces HBM usage while maintaining model convergence accuracy.
  \item Fused Kernels: For example, FlashAttention \cite{dao2023flashattention} and FusedSoftmaxCrossEntropy \cite{ece} combine separate operations into single efficient kernels, reducing launch overhead and improving computational efficiency.
  \item ZeRO \cite{rajbhandari2020zero}: This optimizer shards the dense model state (parameters, gradients, and optimizer states) across GPUs, significantly reducing the HBM consumption on each card and enabling the training of much larger models.
\end{itemize}
\section{Experiment}

\subsection{Benchmark of Operators}
At the operator level, RecIS improves the overall performance of the model through techniques such as operator fusion, vectorized memory access, and atomic operation optimization. Table \ref{tab:MFU} compares the results of key operators with TensorFlow and PyTorch on H20. We use MBU to show the performance of an operator from the start of CPU scheduling to the completion of execution.

\noindent\textbf{Fusion Operations}

\begin{itemize}[leftmargin=*]
  \item Bucketize and Mod: These operations are common feature transformation operators in recommendation system models, often containing dozens or even hundreds of them in a single model. Sequential execution of these operators can result in high GPU scheduling overhead. RecIS supports fusing multiple operators of the same type into a single one, improving GPU efficiency. In our experiments, we simulated 100 columns of features with 10,000 values, performing bucketize or mod operations on each. Although RecIS achieved higher bandwidth utilization with fusion optimization techniques, actual bandwidth utilization remained low. This is due to the additional computation required to fuse the 100 columns of features, a problem that RecIS aims to address in the future.
  \item Ids Partition: In the sparse component, the input IDs are deduplicated and then split according to the sharding strategy. Within the RecIS framework, we implemented an efficient fusion operator to perform these two operations, which requires combining multiple operators in TensorFlow and PyTorch. We used 1 million data for comparative experiments.
\end{itemize}

\noindent\textbf{Reduction Operations}

Reduction is a basic operation for embedding pooling. RecIS provides sum, mean and tile aggregation methods for sparse component. For reduction operations, we used 1 million of 16 dimension data for comparative experiments.
\begin{itemize}[leftmargin=*]
  \item Reduce Sum / Mean: Due to the variable-length nature of actual aggregation inputs, reduction operators can encounter performance issues caused by atomic operations and waste of GPU computing resources due to data sparsity. Therefore, we provide two execution strategies for aggregation operators. When the atomic conflict rate is high, we reduce atomic operations through Warp-level merging combined. When data sparsity is high, we adopt a method that better utilizes GPU resources for computation. We conducted experiments on both high sparsity  (reduce easy) and high atomic conflict (reduce hard) scenarios.
  \item Sequence Tile: RecIS provides a tile reduction operator that supports the concat aggregation method, which can only be implemented in TensorFlow or PyTorch using a combination of the reduce operator and other operators.
\end{itemize}

\noindent\textbf{Embedding Operations}

Embedding operations are also an important part of sparse components. RecIS improves GPU memory bandwidth by vectorized memory access. For embedding operations, we used 1 million of 16 dimension data for comparative experiments.
\begin{itemize}[leftmargin=*]
  \item Gather: Get specified rows from the embedding.
  \item Scatter: Update some rows of the embedding.
\end{itemize}

\begin{table}[ht!]
    \centering
    \caption{Operators' MBU for 3 systems}
    \label{tab:MFU}
    \begin{tabular}{l|
    >{\centering\arraybackslash}m{2cm}| 
    >{\centering\arraybackslash}m{2cm}|
    >{\centering\arraybackslash}m{2cm}}
    \toprule
     & \textbf{TensorFlow} & \textbf{PyTorch} & \textbf{RecIS} \\
    \midrule
    bucketize & 0.40\% & 0.40\% & \textbf{0.88\%} \\
    mod & 0.45\% & 0.70\% & \textbf{1.68\%} \\
    ids partition & - & 34.60\% & \textbf{55.10\%} \\
    sequence tile & 2.43\% & 4.58\% & \textbf{18.25\%} \\
    reduce hard & 0.48\% & 0.93\% & \textbf{2.25\%} \\
    reduce easy & 1.38\% & 2.75\% & \textbf{13.75\%} \\
    gather & 1.70\% & 15.00\% & \textbf{47.50\%} \\
    scatter & - & 20.75\% & \textbf{58.00\%} \\
    \bottomrule
    \end{tabular}
\end{table}

\subsection{End-to-End (E2E) Benchmark}
RecIS has been widely used in Alibaba advertising, recommendation, search, and other scenarios. We chose Model for Search (MSE) and Large Model for Advertisement (LMA) to compare the performance with TensorFlow(with sparse component) and PyTorch(with sparse component). RecIS's core optimizations are mostly in the sparse components, so in addition to comparing the end-to-end model execution time, we also compared the sparse component time.

\subsubsection{MSE}
The primary objective of this model is to accurately predict the probability of a user clicking on a given product within an e-commerce search scenario. The architecture is adept at processing a large set of features, including 13 distinct user historical behavior sequences, by dynamically extracting relevant user interests based on the current search query.

\noindent\textbf{Model Structure}

The model inputs are categorized into two main types: Non-sequential features and sequential features. Represent a total of 660 features for each instance.

The model architecture can be deconstructed into three main stages.
\begin{enumerate}[leftmargin=*]
    \item \textbf{Feature Transformation:} For all sequential and non-sequential features, feature transformation methods can be categorized into three types: \begin{itemize}
        \item Hash: Converts string features to ID type features using hashing (e.g., item names).
        \item Bucketize: Buckets floating-point features according to the bucket index to ID type features (e.g., item prices).
        \item Raw: Floating-point features are directly injected into the model without any transform (e.g., sequence length).
    \end{itemize}
    \item \textbf{Sequential Feature Processing via Cross-Attention:} To dynamically extract user interests from the historical behavior sequence that are relevant to the current search query, the model employs a cross-attention mechanism.
    \item \textbf{Feature Fusion and Deep Network:} The resulting context-aware interest vector is then concatenated with the set of non-sequential features.
    This concatenated result is subsequently fed into a five-layer Deep Neural Network (DNN) to learn high-order non-linear relationships among the features.
    Finally, the output of DNN is passed through a sigmoid activation function to produce the predicted CTR, denoted $pCTR$. This value represents the probability of a click event.
\end{enumerate}

\noindent\textbf{Experiments}

We conducted performance comparison experiments with TensorFlow(with sparse component) and PyTorch(with TorchRec) on 32 H20 GPUs. Under the RecIS framework, the overall runtime was 80\% that of PyTorch and 33\% that of TensorFlow. The sparse computation time was 72\% on PyTorch and 30\% on TensorFlow. MSE has a large number of feature columns, but the data volume of each column is relatively small. This leads to high CPU scheduling overhead and low GPU utilization. Although TensorFlow can leverage graph information on the CPU to alleviate GPU scheduling overhead in a concurrent manner, concurrency is limited on a single machine with multiple GPUs. Furthermore, without operator fusion, each operator cannot fully utilize the GPU's efficiency. RecIS's automatic operator fusion capabilities can combine over 600 feature transformation operators into three operators, effectively reducing GPU scheduling overhead. Furthermore, operator fusion increases the amount of data processed by GPU operators, significantly improving GPU computational efficiency.

\subsubsection{LMA}
The primary objective of this model is to improve the accuracy within an e-commerce advertisement scenario by using lifelong user behavior information.

\noindent\textbf{Model Structure}

The input to the LMA model consists of user behavior sequences encompassing up to 100,000 historical interactions, incorporating both multimodal features (e.g., images and text) and over 400 ID-based categorical features. These features are preprocessed using standard techniques such as sequence truncation, hashing, and feature crossing.

To reduce the computational cost of prediction, LMA compares the target item with 100,000 historical behavior sequences using multimodal similarity matching~\cite{sheng2024enhancing}, and retrieves approximately 100 most relevant past interactions. The features associated with these top-ranked sequences are then used to look up their corresponding embeddings, which are fed into the downstream dense network. The dense component of LMA integrates the ID-based Deep Interest Network (DIN)~\cite{zhou2018deep}, multimodal-based modules including SimTier and MAKE~\cite{sheng2024enhancing}  and other advanced modules
~\cite{sheng2021one,ShengGCYHDJXZ2023JRC,zhou2018deep,BianWRPZXSZCMLX2022CAN}.

\noindent\textbf{Experiments}

We exclusively deployed LMA on TensorFlow and RecIS. This was because the model requires the zero-conflict embedding feature that TorchRec does not currently support. Therefore, we only compared the performance on these two platforms using 64 H20 GPUs.

We used a batch size of 1000 and truncate 16,000 sequential user behavior lifelong features as input in the LMA(16k) experiment. RecIS's overall runtime was 76\% that of TensorFlow, the sparse computation time was 67\% of that of TensorFlow.

We also performed a batch size of 1000 and 100,000 lifelong features of sequential user behavior as input in the LMA(100k) experiment. At this time, TensorFlow cannot work.

\begin{table}[ht!]
    \centering
    \caption{E2E performance for 3 systems}
    \label{tab:E2E}
    \begin{tabular}{l|
    >{\centering\arraybackslash}m{1.5cm}| 
    >{\centering\arraybackslash}m{1.5cm}|
    >{\centering\arraybackslash}m{1.5cm}|
    >{\centering\arraybackslash}m{1.5cm}| 
    >{\centering\arraybackslash}m{1.5cm}|
    >{\centering\arraybackslash}m{1.5cm}}
    \toprule
     & \multicolumn{2}{|c}\textbf{TensorFlow} 
     & \multicolumn{2}{|c}\textbf{PyTorch} 
     & \multicolumn{2}{|c}\textbf{RecIS} \\
     & sparse & overall & sparse & overall & sparse & overall \\
    \midrule
    LMA(16k) & 1137ms & 1374ms & -  & - & \textbf{764ms} & \textbf{1050ms}\\
    LMA(100k) &  -  &  -  & -  & - & \textbf{259ms} & \textbf{588ms}\\
    MSE & 713ms & 1042ms & 303ms & 424ms & \textbf{218ms} & \textbf{337ms}\\
    \bottomrule
    \end{tabular}
\end{table}
\section{Application}

\subsection{Scaling Dense Parameter}
Generative ranking models have demonstrated significant potential in multi-scenario and multimodal content recommendation systems. However, their deployment faces key challenges, including large-scale dense and sparse parameters, high computational complexity, tightly coupled training and inference pipelines, and the need for flexible business extensibility. Conventional frameworks encounter substantial bottlenecks when handling large-scale hash tables, dense computation layers, and joint optimization with reinforcement learning (RL). Specifically, TensorFlow 1.x lacks sufficient flexibility, native PyTorch provides limited support for large-scale sparse parameters, and the TorchRec framework suffers from performance degradation in I/O and multi-machine scalability, further hindered by its closed architecture and slow iteration and open-source cycle.

To address these limitations, we develop a generative recommendation system based on our in-house RecIS framework, which generates ordered user exposure sequences from candidate sets produced by multi-scenario coarse ranking stages. Our generative recommendation system employs a multi-layer attention-based generative architecture that supports flexible auto-regressive decoding strategies and deeply integrates reinforcement learning paradigms into the training pipeline. To handle massive-scale feature storage, RecIS introduces a cross-node distributed storage design and optimization for large hash tables, significantly reducing per-node memory consumption and enabling virtually unlimited feature capacity expansion.

Through systematic efficiency optimizations, RecIS achieves a 200\% increase in maximum batch size per node during training compared to TorchRec, while reducing overall training time by 70\%. Moreover, the system demonstrates near-linear scaling efficiency in large-scale distributed multi-node, multi-GPU settings. Currently, a generative ranking model with 50 million dense parameters has been deployed in A/B testing within the main recommendation feed—a core scenario in our production environment—yielding simultaneous improvements in both user click-through rate and conversion metrics. This real-world deployment validates the effectiveness and scalability of RecIS in industrial-grade recommendation applications.

\subsection{Scaling User Sequence}
Lifelong user behavior modeling is critical for improving the accuracy of recommendation systems. While current state-of-the-art methods can process behavior sequences with lengths up to $10^3$, scaling to longer sequences (up to $10^6$) introduces fundamental challenges in feature i/o,  embedding lookup, and computational compatibility.

\noindent\textbf{Key Challenges}
\begin{itemize}[leftmargin=*]
  \item \textbf{Feature I/O Bottleneck:} 
  Sequences with $10^6$ interactions significantly increase sample size, network bandwidth consumption during training. For example, a single input batch can be as large as 6 GB. Assuming a batch takes 2 seconds to train, the network throughput alone can reach 24 GB/s. Concurrently, due to multiple memory read/write cycles, the peak memory access bandwidth can soar to 100-200 GB/s. These immense bandwidth requirements pose a significant challenge for mainstream hardware. Neither CX7-level network interface bandwidth nor 16-channel DDR memory bandwidth can fully accommodate such a massive data throughput. By using ORC compression and GPU-based sample assembly, we have successfully shifted the training bottleneck away from I/O bandwidth, ensuring highly efficient training..
  \item \textbf{Embedding Lookup Scalability:} 
  Long-term sequences inevitably include massive historical items (e.g., expired products), inflating the embedding table size and degrading lookup efficiency.  For example, for a single batch, the Embedding access volume can reach 1 billion (1B) key-value (KV) queries and 1 terabyte (1TB) of Embedding memory access. By using GPU Unique and GPU Hashtable technologies, we have optimized the Embedding query performance by 10x for each respectively.
  \item \textbf{Tensor Shape Limitation in Legacy Frameworks:} 
  The tensor shape constraint in TensorFlow 1.x ($batchsize \cdot seqlen \cdot embdim \leq 2^{31}-1$) prevents training with ultra-long sequences. But processing large tensors is a standard practice in our new framework. To further optimize this challenge, we leverage mature technologies from the PyTorch ecosystem to save memory and boost efficiency.
\end{itemize}

\noindent\textbf{Business Impact: }
By leveraging RecIS's optimizations for the issues mentioned above, we successfully deployed a user behavior sequence model with a length of 1M in our production environment. This represents a 100-fold improvement over the previous technical standard.

Compared to a 10k sequence length, the model achieved a 4.8\% increase in CTR while simultaneously reducing training costs by 50\%.

\subsection{Scaling Modality}
Unlike conventional approaches that scale recommendation systems by simply increasing dense parameters or sequence length, we explore an alternative dimension of scaling: scaling modality. This paradigm shifts the focus from model size to functional expressiveness—enabling rich multi-modal understanding through the integration of large general-purpose foundation models with compact, task-specific expert modules.

This hybrid architecture decouples generic knowledge acquisition from domain-specific adaptation, effectively addressing operational constraints while maintaining computational feasibility in high-throughput, real-time serving environments. A key enabler of this framework is target-aware embeddings, a knowledge transfer mechanism that aligns general representations with downstream tasks through contextual modulation.

In practice, such foundation models must jointly capture intra-modal semantics and inter-modal interactions—posing two core challenges:
\begin{itemize}[leftmargin=*]
  \item How to unify and efficiently represent diverse modalities—including large-scale sparse ID embeddings and dense semantic features—within a single coherent framework;
  \item How to efficiently fuse heterogeneous embeddings via dense computation without sacrificing scalability or performance.
\end{itemize}

Leveraging the RecIS framework—a system co-designed for large-scale sparse-dense computation—we validate this architectural paradigm from both user-centric and item-oriented perspectives. Comprehensive offline evaluations and online A/B tests across multiple industrial recommendation scenarios demonstrate consistent and significant improvements in key metrics, confirming the effectiveness of modality scaling powered by RecIS.
\section{Related Work}
\noindent\textbf{Modeling: }
A standard DLRM framework integrates diverse input signals. Two of them are behavioral sequence  and feature interaction. \cite{pi2020search,si2024twin,zhou2018deep} leverages target-aware attention to quantify correlations between user historical actions and candidate items. \cite{lian2018xdeepfm,tang2020progressive,wang2021dcn,wang2025home,BianWRPZXSZCMLX2022CAN} captures high-order relationships across heterogeneous features to generate final prediction scores.
\cite{zhang2024wukong, guo2023embedding}  scaling paradigm emphasizes expanding the feature interaction component – the module responsible for fusing user-item representations. \cite{han2024enhancing,zhang2024scaling,shin2023scaling,yan2025unlocking}  prioritizes user tower expansion, where only user-side components undergo parameter scaling, yielding inference-optimized architectures. Alternative approaches like GRM demonstrate complementary scaling capabilities: \cite{zhai2024actions} validate scaling laws through HSTU models with trillion-parameter capacity, while \cite{deng2025onerec} replace conventional ID embeddings with semantic coding, integrating DPO optimization within transformer architectures to establish a unified generative framework replacing traditional cascaded pipelines.

\noindent\textbf{System: }
The industry has seen many optimizations based on general frameworks. \cite{TFRS,DeepRec,zhang2022picasso} enhances the sparse computation capabilities of TensorFlow; NVIDIA provides \cite{oldridge2020merlin,wang2022merlin} as a prime example of taking advantage of GPUs for recommendation models to the extreme. 
\cite{torchrec} is a plugin that supports Table Batched Embedding (TBE) in the PyTorch ecosystem. \cite{recsys-examples} implemented dynamic embeddings.
\cite{torcheasyrec,DeepCtr-Troch} are plugins for algorithmic model structures.

\section{Conclusion}
In this report, we introduce RecIS, a PyTorch-based unified sparse-dense training framework for recommendation models. RecIS ported the sparse components from the TensorFlow based training framework. In addition, RecIS is holistically optimized for IO, memory access, and computing to achieve efficient computation of sparse and dense components, achieving up to \textbf{2$\times$ higher} training throughput. Currently, RecIS is being used in Alibaba for numerous large-model enhanced recommendation
training tasks.

\clearpage
\section{Authors}

Within each role, authors are listed alphabetically.

\begin{multicols}{2}


\vspace{1em}
\textbf{\textcolor{orange!80!black}{Contributors}}\\
\begin{itemize}[leftmargin=1.5em, itemsep=1pt]
    \item Hua Zong
    \item Qingtao Zeng
    \item Zhengxiong Zhou
    \item Zhihua Han
    \item Zhensong Yan
    \item Mingjie Liu
    \item Hechen Sun
    \item Jiawei Liu 
    \item Yiwen Hu
    \item Qi Wang
    \item YiHan Xian
    \item Wenjie Guo
    \item Houyuan Xiang
    \item Zhiyuan Zeng
    \item Xiangrong Sheng
    \item Bencheng Yan
    \item Nan Hu
    \item Yuheng Huang
    \item Jinqing Lian
    \item Ziru Xu
    \item Yan Zhang
    \item Ju Huang
    \item Siran Yang
    \item Huimin Yi
\end{itemize}

\vspace{1em}
\textbf{\textcolor{orange!80!black}{Supervision}}\\
\begin{itemize}[leftmargin=1.5em, itemsep=1pt]
    \item Jiamang Wang
    \item Pengjie Wang
    \item Han Zhu
    \item Jian Wu
    \item Dan Ou
    \item Jian Xu
    \item Haihong Tang
    \item Yuning Jiang
    \item Bo Zheng
    \item Lin Qu
\end{itemize}

\end{multicols}


\clearpage
\bibliography{ref}
\bibliographystyle{colm2024_conference}

\end{document}